\documentclass[proceedings, preprint]{rmaa}

\usepackage{paralist}

\usepackage{psfrag,color}

\usepackage{multicol}
\usepackage{multirow}

\usepackage{amsmath}
\usepackage{graphicx}
\usepackage{breqn} 
\usepackage[mathscr]{euscript} 

\SetYear{2022}
\SetConfTitle{Prospects for low-frequency radio astronomy in S. A.}

\title{Super winds and radio emission in X-ray binary systems} 

\author{
  L. Abaroa\altaffilmark{1,2} 
  and G.E. Romero\altaffilmark{1,2}}

\altaffiltext{1}{Instituto Argentino de Radioastronomía, CICPBA-CONICET-UNLP, Villa Elisa, La Plata, Argentina.}

\altaffiltext{2}{Facultad de Cs. Astron\'omicas y Geof\'{\i}sicas, Universidad Nacional de La Plata, Paseo del Bosque S/N (1900), La Plata, Argentina.}

\shortauthor{Abaroa \& Romero}
\shorttitle{Super winds and Radio emission in XRBs}

\listofauthors{L. Abaroa, \& G.E. Romero}
\indexauthor{Abaroa, L.}
\indexauthor{Romero, G.E.}

\abstract{We have recently proposed that supercritical colliding wind binaries (SCWBs) are suitable scenarios for particle acceleration and nonthermal radiation. In these X-ray binary systems (XRBs), the wind from the companion star collides with the wind ejected from the super-Eddington accretion disk of the stellar black hole. Strong shocks are generated in this collision, leading to the acceleration of particles and subsequent broadband emission through different nonthermal radiative processes. In particular, we estimate luminosities of the order of $L\approx 10^{34}\,{\rm erg\,s^{-1}}$ in the radio band. One of the major components in these processes is the power provided by the super wind expelled from the disk. Furthermore, some properties of the wind photosphere, such as its geometry or its temperature distribution, also contribute to the absorption and reprocessing of the nonthermal radiation. In this work, we perform a more detailed description of the powerful wind launched from the accretion disk, in order to obtain a better understanding of the above-mentioned processes.}

\resumen{Hemos propuesto recientemente que las binarias con colisión de viento súper críticas (SCWBs) pueden ser escenarios apropiados para acelerar partículas y producir radiación no térmica. En estas binarias de rayos X (XRBs), el viento de la estrella donante de masa colisiona con el viento eyectado del disco súper crítico del agujero negro estelar. Como consecuencia de esta colisión se generan ondas de choque, en donde se aceleran partículas y se produce radiación de origen no térmico a través de distintos procesos radiativos. En particular, estimamos que, en la banda de radio del espectro, la luminosidad puede alcanzar valores de $L\approx 10^{34}\,{\rm erg\,s^{-1}}$. El principal componente de estos procesos es el súper viento expulsado del disco de acreción. Además, algunas propiedades de la fotosfera asociada al viento, tales como su geometría o distribución de temperaturas, también contribuyen en la posterior absorción y reprocesamiento de la radiación no térmica. En este trabajo, describimos de manera más detallada el viento  lanzado del disco de acreción, con el propósito de tener un mejor entendimiento de los procesos mencionados.}

\addkeyword{X-ray binaries}
\addkeyword{outflows}
\addkeyword{black holes}
\addkeyword{accretion disks}

\begin{document}
\maketitle

\section{Introduction}
\label{sec:intro}
Supercritical colliding wind binaries (SCWBs) have recently been proposed as suitable scenarios for particle acceleration and nonthermal emission \citep{Abaroa2023}. These systems are X-ray binaries composed of a star and a stellar black hole accreting at super-Eddington rates. At these accretion rates, the radiation pressure on the surface of the disk overcomes the gravitational attraction, and matter is expelled in the form of a strong wind. If the donor star is a hot early-type star, its wind can collide with that of the accretion disk, giving rise to a SCWB. Such systems have strong shocks and are particle accelerators and nonthermal emitters \citep{Abaroa2022,Abaroa2023}. 

The collision of winds in SCWBs is similar to that of colliding wind binaries (CWBs), in which the powerful winds from two massive stars collide. Contrary to the SCWBs, these systems are widely studied through both, observational and theoretical tools \citep{Benaglia2003,De2013,Benaglia2015,Pittard2021}. The most luminous radio-emitting CWB is WR140, with a total radio luminosity of $\sim 2.6\times 10^{30} \, \rm{erg \, s^{-1}}$, whereas \citet{Abaroa2023} showed that SCWBs from nearby galaxies can produce radio luminosities of the order of $\sim10^{34}\,{\rm erg\,s^{-1}}$. However, few X-ray binaries are identified as supercritical, and most of them are extragalactic sources, while there are tens of identified CWBs \citep{De2013}. Another important difference between CWBs and SCWBs is that the former has a relatively faint emission at high energies (with only two cases confirmed, $\eta$ Carinae and WR11), whereas the latter should exhibit gamma-ray emission of $\sim10^{34}\,{\rm erg\,s^{-1}}$. 

We also showed that the radio luminosity of SCBWs is typically a fraction $\sim 10^{-5}$ of the power of the disk-driven wind, so the investigation of the wind expelled from the accretion disk is essential to understand the radio luminosity of these sources.

In this work, we develop a more detailed  description of the wind launched from the surface of the accretion disk of the black hole in SCBWs. A better understanding of this wind can help to improve the investigation of the subsequent collision of winds in SCBWs. We start, in $\S$2, by describing the model we use to investigate the properties of the disk-driven wind. In $\S$3 we show the results and in $\S$4 we close with a summary and conclusions.

\section{Model}

We assume that the X-ray binary is composed of a Population I star and a non-rotating stellar mass black hole (BH) in a close orbit. 

The orbital semi-axis $a$, the stellar radius, and the mass ratio of the system, $q=M_*/M_{\rm BH}$, satisfy \citep{Egleton1983}:
\begin{equation} \label{eggleton}
    R_{\rm{lob}}^*=  \dfrac{a \ 0.49 \ q^{2/3}}{0.6 \ q^{2/3} + \ln{(1+q^{1/3})}}\mathbf{,}
\end{equation}
where $M_*$ is the mass of the star and $M_{\rm BH}$ the mass of the BH.
Hence, the star overflows its Roche lobe $R_{\rm{lob}}^*$, transfers mass to the BH through the Lagrange point, and an accretion disk is formed due to the angular momentum of the system. 


We assume a Newtonian potential for the gravity field, since we are interested in weak-field processes.

\subsection{Accretion disk} \label{subsec: accretion disk}

We adopt cylindrical coordinates with axial symmetry along the $z$-axis, neglect the self-gravity of the disk, and consider a non-magnetized disk with a super-Eddington accretion rate at the outer part, $\dot{m}_{\rm input}=\dot{M}_{\rm input}/\dot{M}_{\rm Edd} \gg 1$, where $\dot{M}_{\rm input}$ is the input mass per time unit in the accretion disk. 
The Eddington rate is given by
\begin{equation} \label{tasa critica}
    \Dot{M}_{\rm{Edd}}=   \frac{L_{\rm{Edd}}}{\eta c^2} \approx 2.2\times 10^{-8} M_{\rm BH} \ {\rm yr^{-1}},
\end{equation}
with $L_{\rm Edd}$ the Eddington luminosity\footnote{The Eddington luminosity is defined as the luminosity required to balance the attractive gravitational pull of the accreting object by radiation pressure.}, $\eta \approx 0.1$ the accretion efficiency, $c$ the speed of light, and $M_{\rm BH}$ the black hole mass.

The critical radius, given by
\begin{equation}
    r_{\rm crit} \sim  40\,\dot{m}_{\rm input}\, r_{\rm g},\label{eq:rg}
\end{equation}
separates the disk in two regions: a standard outer disk \citep{1973A&A....24..337S} and a radiation-dominated inner disk with advection \citep{Fukue2004}. In relation (\ref{eq:rg}), $r_{\rm g}=GM_{\rm BH}/c^2$ is the gravitational radius of the BH, with $G$ the gravitational constant.

In the disk model the advection is parameterized as a fraction $f$ of the viscous heating, $Q_{\rm adv}=fQ_{\rm vis}$. The disk becomes geometrically thick in the inner region, where the ejection of winds by the radiation force helps to regulate the mass-accretion rate onto the BH ($\dot{M}_{\rm acc}$) at the Eddington rate\footnote{$\dot{M}_{\rm acc}=\dot{M}_{\rm input}$ in the outer region of the disk and $\dot{M}_{\rm acc}=\dot{M}_{\rm input}r_{\rm d}/r_{\rm crit}$ in the inner region \citep{Fukue2004}.}. 

Because of the high opacity, the disk is optically thick. We assume that it radiates locally as a blackbody. The radiation intensity  of a plasma element in the comoving frame with the outer and inner parts of the disk, at a radius $r_{\rm d}$ measured on the equatorial plane, is given by:

\begin{equation}\label{intensidad}
     I_0=\frac{1}{\pi}\sigma T_{\rm eff}^4 =
    \left\lbrace \begin{array}{l}
    \dfrac{1}{\pi}\dfrac{3GM_{\rm BH}\dot{M}_{\rm input}}{8\pi r_{\rm d}^3} f_{\rm in}, \ \ r_{\rm d} > r_{\rm crit}\\ \\
  \dfrac{1}{\pi}\dfrac{3}{4}\sqrt{c_3}\dfrac{L_{\rm Edd}}{4\pi r_{\rm d}^2}, \ \ r_{\rm d} \le r_{\rm crit},
    \end{array} 
    \right.
\end{equation}
where $\sqrt{c_3}=H/r_{\rm d}=\tan{\delta}$, with $H$ the disk height scale, $\delta$ the disk opening angle, and $f_{\rm in}=1-r_{\rm in}/r_{\rm d}\approx1$ \ (as $r_{\rm d}>r_{\rm crit}$, then $r_{\rm d}\gg r_{\rm in})$. Here, $c_3$ (as well as $c_1$ and $c_2$ used in the next section) is a coefficient that depends on the advection parameter, the adiabatic index of the gas $\gamma$, and the viscosity $\alpha$ \citep[see Appendix in][]{Fukue2004}. We adopt a disk with $f=0.5$ and $\alpha=0.5$; that means, we assume equipartition between advection and viscous heating. The index $\gamma=4/3$ corresponds to a radiation-dominated gas in the inner disk. These values lead to a disk-opening angle of $\delta=30^{\circ}$.

\subsection{Radiation fields} \label{subsec: radiation fields}

In order to obtain the radiative contribution of each plasma element,  $\mathscr{Q}=(r_d,\phi_d,H)$, of the disk surface, at any point, $\mathscr{P}=(r,\phi,z)$, above or below the disk, we make a transformation of the intensity between the inertial and comoving reference frames. Azimuthal symmetry allows us to perform the calculations for any constant value of $\phi$; therefore, we do it in the $rz$ plane $(\phi=0)$.

The relativistic Doppler factor $\mathscr{D}$ provides the transformation between the reference frames \citep{Mckinley1980}:
\begin{equation} \label{intensidad}
    I=\mathscr{D}^4 I_0=\frac{I_0}{(1+z_{\rm red})^4},
\end{equation}
where $z_{\rm red}$ is the redshift factor given by \citep{1999PASJ...51..725W}
\begin{equation}
    z_{\rm red}=-\frac{(r \cos{\phi_{\rm d}} - r_{\rm d})v_r - (r \sin{\phi_{\rm d}}) v_\phi + (z-H)v_r c_3}{cD}.
\end{equation}
In this expression, $D$ is the distance between $\mathscr{P}$ and $\mathscr{Q}$, $v_\phi=c_2 v_{\rm K}$ is the azimuthal velocity and $v_r=-c_1 \alpha v_{\rm K}$ is the radial velocity, with $v_{\rm K}=\sqrt{GM_{\rm BH}/r_{\rm d}}$ the Keplerian velocity. We want to notice that we only consider the inner part of the disk for these calculations, because the intensity decays with $r_{\rm d}^{-3}$.

The radiation-field tensor is given by \citep{Rybicki1986}
\begin{equation} \label{eq: radiation tensor}
    R^{\mu \nu}=\begin{pmatrix}
    E & \frac{1}{c}F^{\alpha} \\ 
    \frac{1}{c}F^{\alpha} & P^{\alpha \beta}
    \end{pmatrix}= \frac{1}{c}\int Ij^{\mu}j^{\nu} \rm{d}\Omega
\end{equation}
where $I$ is the intensity given by Eq. \ref{intensidad}. 
This is a symmetric tensor of rank $2$ and therefore we calculate ten elements in total: one for the energy density $E$, three for the flux vector $F^{\alpha}$, and six for the stress tensor $P^{\alpha \beta}$. In Eq. \ref{eq: radiation tensor}, $j^{\mu}$ and $j^{\nu}$ are the direction cosines in Cartesian coordinates, and $\Omega$ is the solid angle subtended by $\mathscr{Q}$.

\subsection{Particles in the photon field} \label{sect: wind of the disk}

We now calculate the trajectory and velocity of the particles ejected from the disk when they interact with photons of the radiation field. The equation of motion under a relativistic, radiation treatment, is given by \citep{2020fafd.book.....K}:
\begin{equation}\label{conservation}
    f_{\mu} = -\frac{\partial \Phi_{\rm e}}{\partial x^{\nu}} + R_{\mu;\nu}^{\nu},
\end{equation}
where $f_{\mu}$ is the four-force per unit volume. The effective potential $\Phi_{\rm e}$ is the sum of gravitational $(\Phi_{\rm g})$ and centrifugal $(\Phi_{\rm c})$ potentials. The semicolon $(;)$ in the second term refers to the covariant differentiation of the energy-momentum tensor. 

As we consider a disk with axial symmetry, the gravitational potential cancels out in the azimuthal coordinate: $\partial \Phi_{\rm g}/\partial x^{\alpha}=(\partial \Phi_{\rm g}/\partial r,0,\partial \Phi_{\rm g}/\partial z)$.
Furthermore, the centrifugal potential acts only in the radial direction: $\partial \Phi_{\rm c}/\partial x^{\alpha}=(l^2/r^3,0,0)$, with $l=r_{\rm d}^2\omega_{\rm K}$ being the specific angular momentum of the disk, and  $\omega_{\rm K}$ the angular velocity.

The equations of motion of the ejected particles can be found working with Eq. (\ref{conservation}) \citep[see Eqs. $11-13$ in][]{Abaroa2023}. We solve these equations  numerically and assume that the kinematics of the disk-driven wind is roughly described by the trajectory and terminal velocities obtained for the test particles. As the accretion rate in the inner region of the disk is regulated at the Eddington rate, the mass loss in the wind is of the order of the super-Eddington accretion rate, $\dot{M}_{\rm dw}\sim \dot{M}_{\rm input}$.

\subsection{Wind emission}

We calculate the thermal emission of the photosphere of the disk-driven wind assuming a spherically symmetric wind that expands with constant velocity equal to its terminal velocity. Since the mass-loss rate of the disk is much higher than the critical rate, the wind is optically thick and therefore we assume that it radiates locally as a blackbody. The temperature measured by an observer at infinity is given by \citep{2009PASJ...61.1305F}:
\begin{equation}
     \sigma_{\rm T} T_{\rm ph}^4=\frac{\dot{e} \, L_{\rm Edd}}{(1-\beta \cos{\Theta})^4 \, 4 \pi R^2}
,\end{equation}
where $\dot{e}=\dot{E}/L_{\rm Edd}$ is the normalized comoving luminosity, $\beta=v_{\rm w}/c$ the normalized velocity, $\Theta$ the angle of the flow with respect to the line of sight, and $R=\sqrt{r^2+z^2}$, with $r$ and $z$ the being cylindrical coordinates. We assume that the comoving luminosity is equal to the Eddington luminosity ($\dot{e}=1$), as is commonly done in supercritical wind-models \citep[e.g.,][]{2009PASJ...61.1305F}.

The apparent photosphere of this wind is defined as the surface where the optical depth $\tau_{\rm photo}$ is equal to 1 for an observer at infinity. If the velocity of the wind is relativistic, the optical depth in the observer frame depends in general on the magnitude of the velocity and the viewing angle. The location of the apparent photosphere from the equatorial plane $z_{\rm photo}$ is:
\begin{equation}
    \tau_{\rm ph}=\int^\infty_{z_{\rm ph}} \gamma_{\rm w}(1-\beta \cos{\Theta}) \, \kappa_{\rm co} \,\rho_{\rm co} {\rm d}z =1,
\end{equation}
where $\gamma_{\rm w}$ is the wind Lorentz factor, $\kappa_{\rm co}$ the opacity in the comoving frame, and $\rho_{\rm co}$ the wind density in the comoving frame. As we assume a fully ionized wind, the opacity is dominated by free electron scattering ($\kappa_{\rm co}=\sigma_{\rm T}/m_{\rm p}$).

\section{Results}

In this section, we present the results we obtained for the radiation field produced by the accretion disk and the ejected wind, following the model described in the previous section. We apply our model to a black hole and accretion disk with the parameters shown in Table \ref{tab}.

\begin{table}[ht]
\begin{center}
\begin{tabular}{l c c}
\hline
\rule{0pt}{2.5ex}Parameter & Value & Units \\  
\hline
\rule{0pt}{2.5ex}$M_{\rm{BH}}$    & 10 & $M_{\odot}$ \\
$\dot{M}_{\rm input}$   & $10^2$  & $\dot{M}_{\rm Edd}$ \\
$r_{\rm g}$    & $1.5\times10^6$ & cm  \\
$\dot{M}_{\rm Edd}$   & $2.2\times10^{-7}$ & $M_{\odot}\ \rm yr^{-1}$\\
$v_{\rm w}$ & $5\times 10^{9}$ & cm $\rm s^{-1}$ \\
$L_{\rm kin}^{\rm w}$ & $1.5\times 10^{39}$ & erg $\rm s^{-1}$ \\
$\alpha$ & $0.5$ & $-$ \\
$f$ & $0.5$ & $-$ \\
$\gamma$ & $4/3$ & $-$ \\
\hline
\end{tabular}
\end{center}
\caption{Parameters adopted in our model for the black hole, the accretion disk, and its wind.
} 
\label{tab}
\end{table}

\subsection{Radiation field}

\begin{figure*}[ht]
\begin{multicols}{3}
    \includegraphics[width=\linewidth]{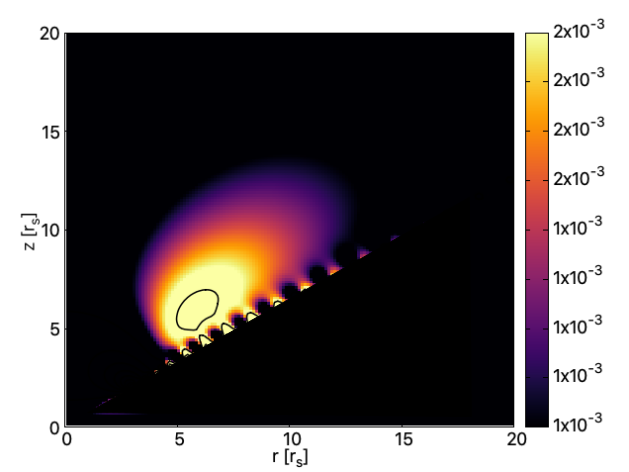}\par 
    \includegraphics[width=\linewidth]{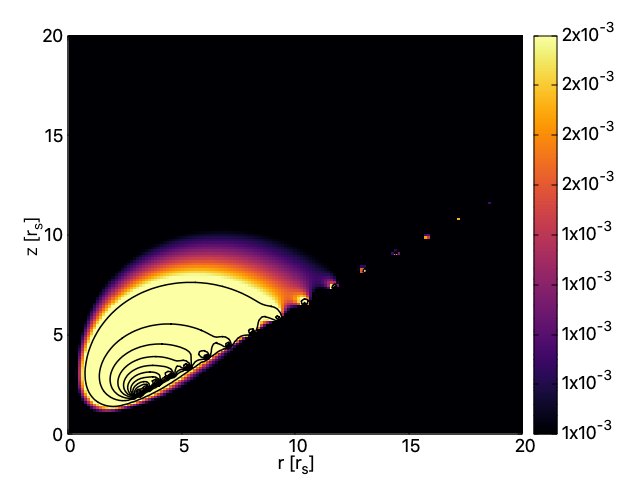}\par 
    \includegraphics[width=\linewidth]{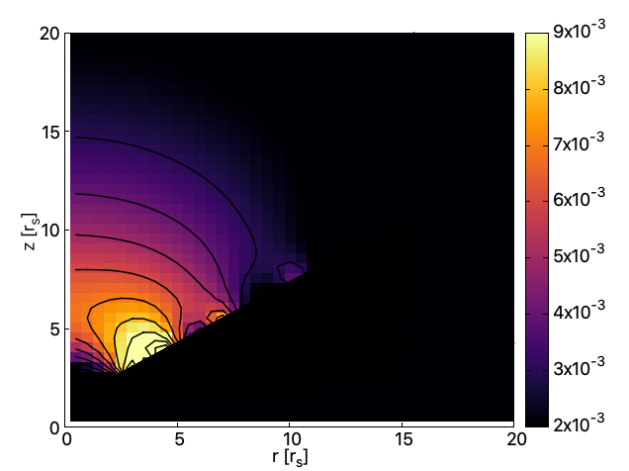}\par     
    \end{multicols}
\caption{Cylindrical components of the flux vector of the radiation field tensor. The vertical axis is the $z$-distance and the horizontal axis is the $r$-distance from the black hole, measured in units of Schwarzschild radius. From left to right: radial, azimuthal, and vertical components. The color bar gives the normalized value of each flux component. Small loops at the surface of the accretion disk are due to numerical fluctuations and have no physical meaning.}
\label{fig: radiation fields}
\end{figure*}

We solve Eq. (\ref{eq: radiation tensor}) to find the elements of the radiation-field tensor, that is, the energy density, flux, and pressure produced by the accretion disk. Fig. \ref{fig: radiation fields} shows the spatial distribution normalized components of the radiation-flux vector (in cylindrical coordinates). Horizontal and vertical axis are coordinates in a meridional plane, in units of Schwarzschild radius. We note that, in the innermost part of the disk, the azimuthal component of the flux is far more relevant than the radial and vertical components. This is mainly because of the semi-relativistic angular velocity of the accretion disk in the vicinity of the black hole, where some radiation is blueshifted and some is redshifted, resulting in differential azimuthal fluxes. 

\subsection{Wind photosphere}

\begin{figure}[ht] 
        \centering          \includegraphics[width=\columnwidth]{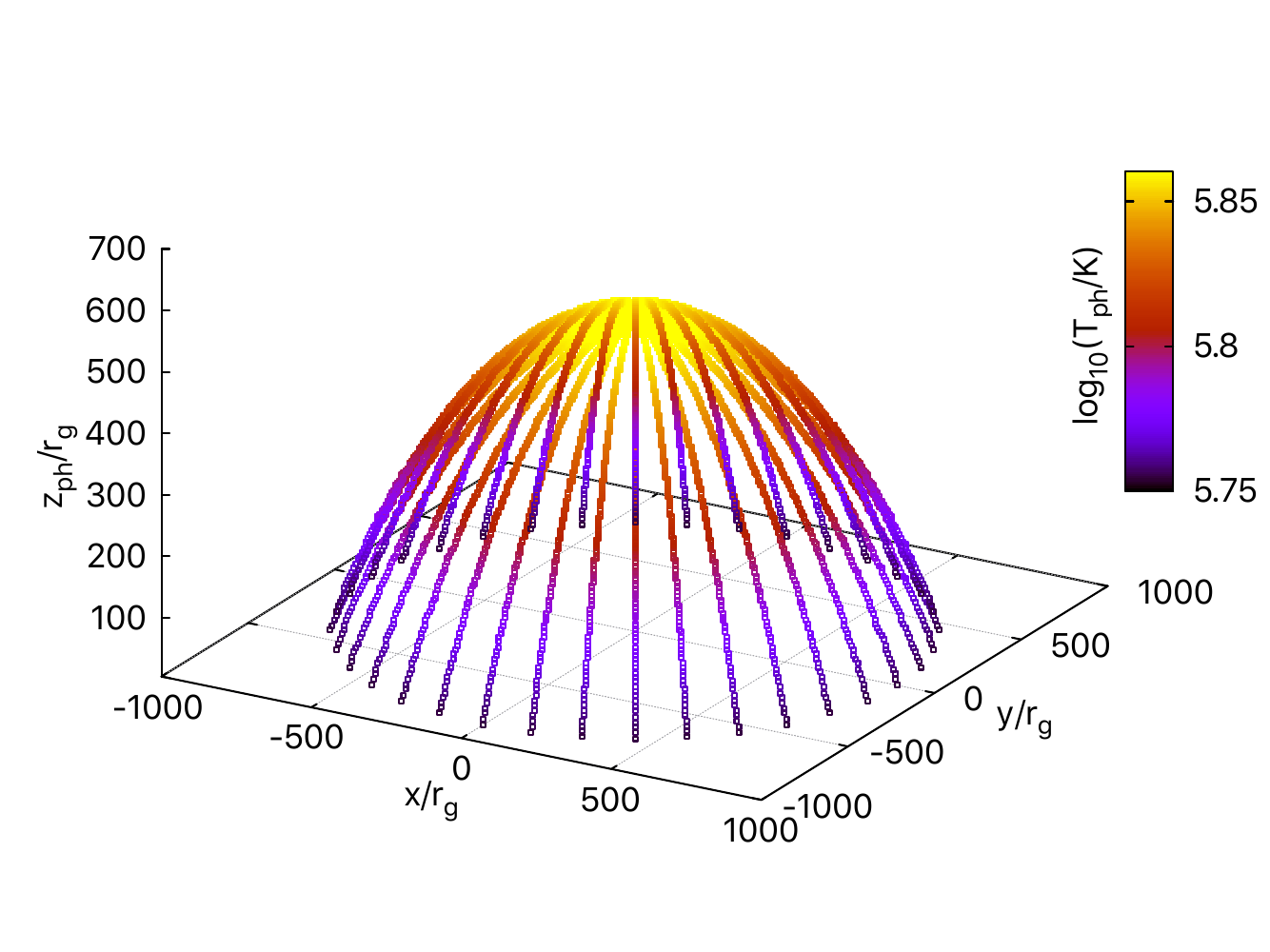}
            \caption{Wind photosphere in three dimensions (Cartesian coordinates in units of gravitational radius). The photosphere has a spheroidal shape with a height of $z_{\rm ph}\sim600\,r_{\rm g}$, and a mean temperature of $T\sim5\times 10^5{\rm K}$. The top of the photosphere is not sunk because of the low value of the Lorentz factor.}
\label{fig: photosphere}
\end{figure}

\begin{figure}[ht] 
        \centering          \includegraphics[width=\columnwidth]{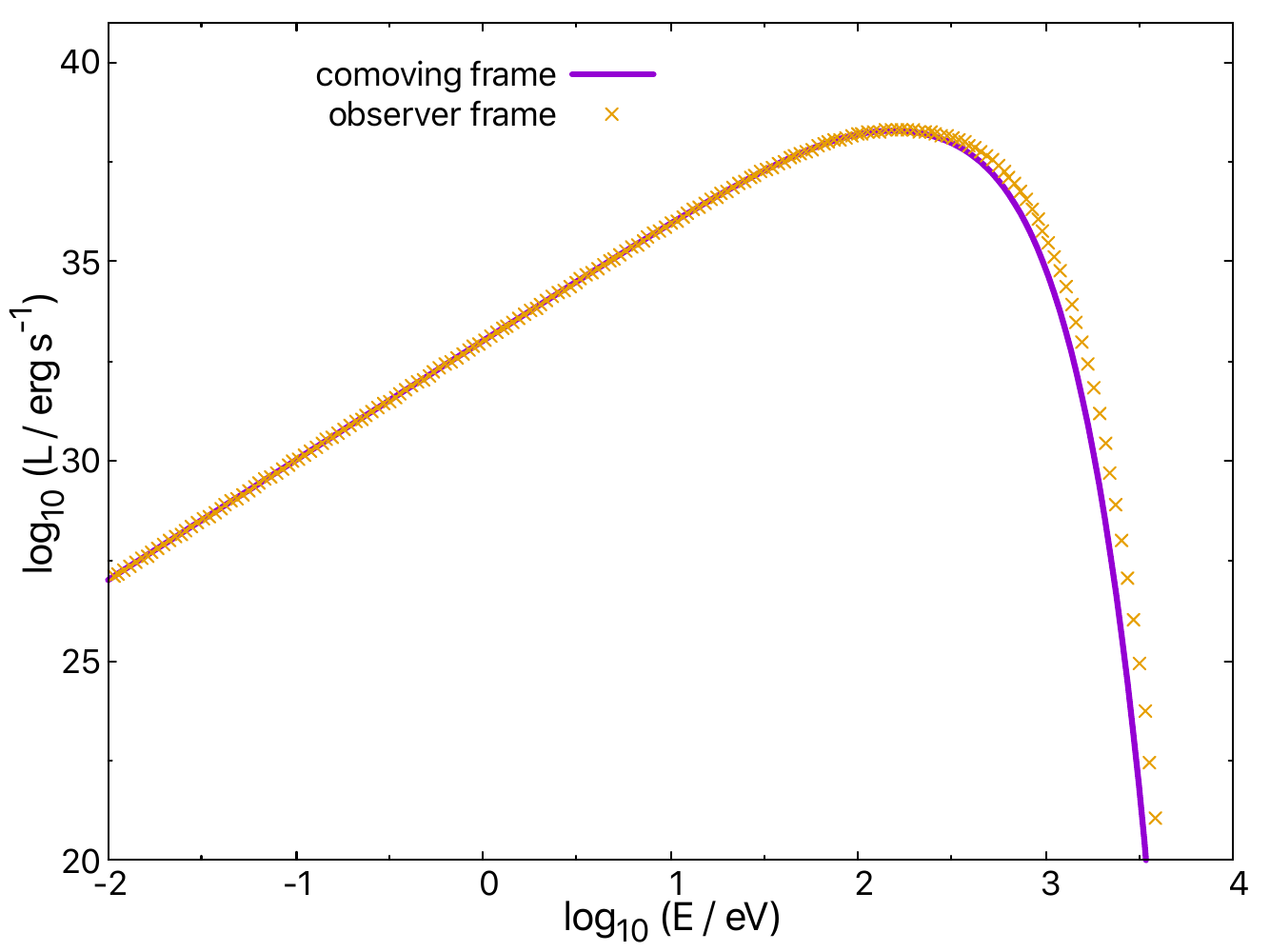}
            \caption{Spectrum of the radiation produced by the wind photosphere in the comoving reference frame (solid line) and the observer frame (dots). Both curves are almost equal, because of the low value of the Lorentz factor.}
\label{fig: spectrum}
\end{figure}

Fig. \ref{fig: photosphere} shows the geometry of the wind photosphere and its comoving temperature (color bar). The photosphere has a spheroidal shape, with a height of $z_{\rm ph}\sim600\,r_{\rm g}$ and a temperature of the order of $T_{\rm ph}\sim5\times10^5\,{\rm K}$. We also calculate the spectrum of the photosphere, in the observer and comoving reference frames (Fig. \ref{fig: spectrum}). We note that there are no significant differences between both reference frames. This is because the velocity of the wind, despite being semi-relativistic, does not have a Lorentz factor high enough for the relativistic effects to be relevant. This is also why the photosphere has a spheroidal shape and it is not sunk at the top \citep[compare with, e.g., Figure 1 from][]{2009PASJ...61.1305F}.

\subsection{Particles in the wind}

\begin{figure*}[ht]
\begin{multicols}{3}
    \includegraphics[width=\linewidth]{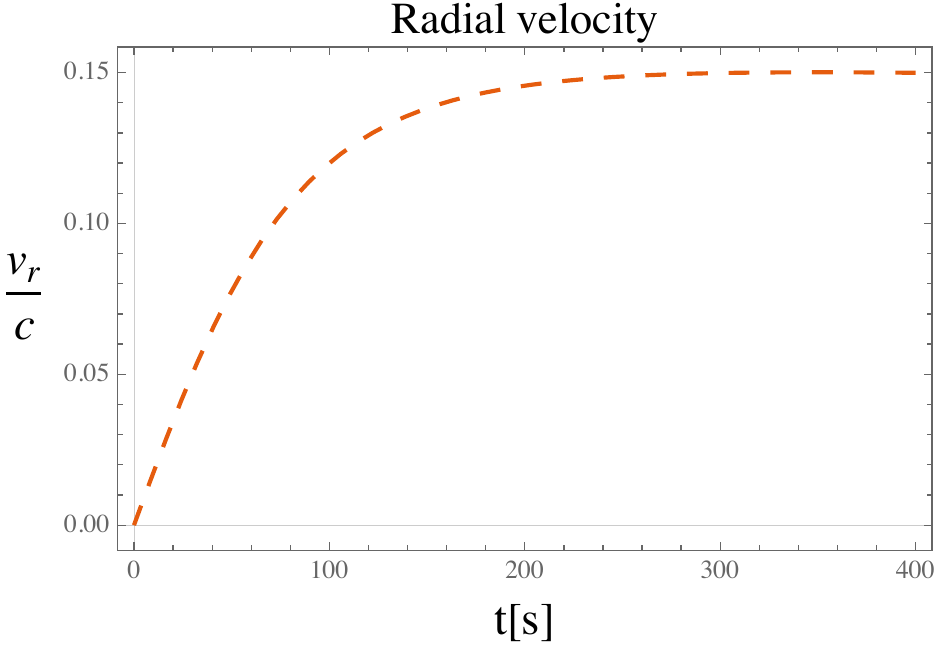}\par 
    \includegraphics[width=\linewidth]{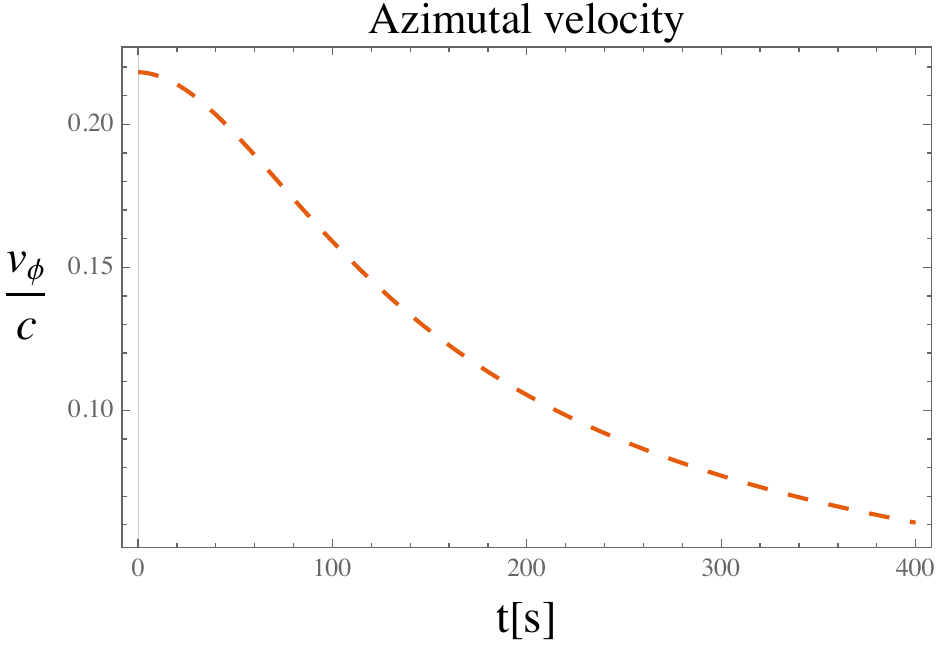}\par 
    \includegraphics[width=\linewidth]{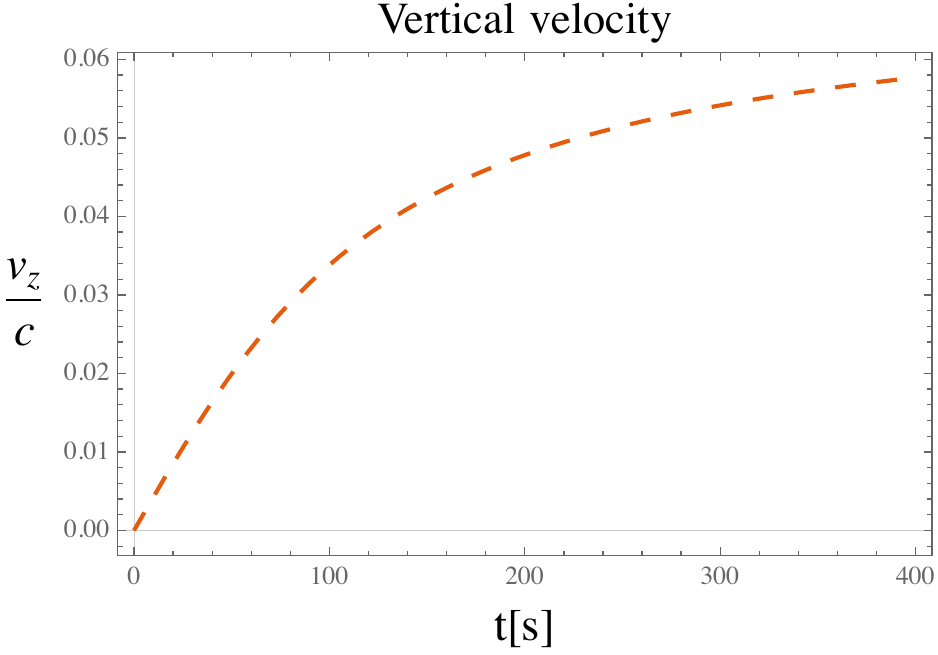}\par     
    \end{multicols}
\caption{Cylindrical components of the velocity for a test particle launched at a radius $r_0=40\,r_{\rm g}$. Since the angular velocity in the vicinity of the black hole is semi-relativistic, we set as initial conditions the radial and vertical components to zero. The vertical axis is the velocity in units of the speed of light, and the horizontal axis is the time in seconds.}
\label{fig: velocities}
\end{figure*}

The initial condition imposed on the particles ejected from the accretion disk is that they only have an azimuthal component of the velocity. Initial radial and vertical velocities are assumed to be zero, because on the surface of the disk $v_{\phi}\gg v_{r},v_{z}$. The wind launched from the radiation-dominated region of the disk will be determined by the radiation forces acting on the particles on the disk surface and along their subsequent trajectories. These forces will have contributions from different parts of the disk in relative motion with respect to the particles. Fig. \ref{fig: velocities} shows the components of the velocity for a test particle launched from a distance $r=40r_{\rm g}$ of the black hole. The azimuthal component reaches its maximum when launched, but soon tends to zero. The radial velocity reaches a constant value of $v_{r}\sim45.000~{\rm km\,s^{-1}}$, and the vertical $v_{z}\sim 18.000~{\rm km\,s^{-1}}$. As seen from the figures, the radial component will dominate the trajectory of the test particle at infinity. The stationary state is reached when the intensity of the radiation field becomes negligible, at a distance of $\sim200r_{\rm g}$ from the black hole. From there on, the ejected particles can be treated as a gas outflow with two components: an equatorial wind with a velocity of $\sim 4.5\times 10^4$ km s$^{-1}$, and a bipolar outflow moving at $\sim 2\times 10^4$ km s$^{-1}$.


\section{Summary and Conclusions}
Moderate supercritical black holes can supply enough power to launch very powerful winds that will reach the donor star in X-ray binaries. These dense and semi-relativistic winds will collide with those of the star, where shocks should form. Particles can be accelerated in these shocks and produce nonthermal radiation. Synchrotron radio emission is of particular interest because should be detected in some nearby sources with ALMA in the sub-mm range or with VLA at longer wavelengths. 

Hypercritical sources ($10^{3-4}\dot{M}_{\rm Edd}$) could display radio luminosities higher than those found by \citet{Abaroa2023}. The energy available to produce such luminosities is primarily provided by the kinetic luminosity of the disk-driven wind. Then, a proper investigation of this wind is necessary to shed light on the transfer of the power of the wind to relativistic particles. Two improvements that can help to this goal are: i) taking into account general relativistic effects, for a more accurate description of the particles ejected from the innermost part of the accretion disk, and ii) making a full fluid treatment of the launching and propagation of the disk-driven wind. 

Nevertheless, we note that a simplified and conservative wind model like the one we apply in this work is suitable for the prediction of the order of magnitude of the radio luminosity of these peculiar systems. These estimates show that their detection is feasible. 

In future works, we will explore the contribution of these systems to the population of cosmic rays in the host galaxy. Most of the protons accelerated in the colliding wind region will be advected away without significant energy losses. Such protons will be transported to the region where they will interact with the interstellar medium. There they might be re-accelerated producing a source of cosmic rays and perhaps a gamma-ray halo through $pp\rightarrow pp+\pi^0$ interactions. Re-accelerated electron and secondary pair from the hadronic interactions can also produce a radio halo surrounding these sources, as observed, for instance, in the extragalactic microquasar S26.



\end{document}